\documentclass[a4paper,twocolumn,english,aps,accepted=2023-02-07]{quantumarticle}
\pdfoutput=1

\usepackage[utf8]{inputenc}
\usepackage{amsmath,amssymb,bbm,mathrsfs,bm,braket,mathtools,color,comment}
\usepackage{graphicx}
\usepackage{appendix}
\usepackage{hyperref}
\usepackage[mathscr]{euscript}
\usepackage{cite}
\usepackage[labelfont=bf]{caption}
\usepackage{float}
\newcommand{\cH}{\mathcal{H}}
\newcommand{\cP}{\mathcal{P}}
\newcommand{\cL}{\mathcal{L}}

\begin{document}

\title{Parity Quantum Optimization:~Compiler}
\author{Kilian Ender}
\affiliation{Parity Quantum Computing GmbH, A-6020 Innsbruck, Austria}
\affiliation{Institute for Theoretical Physics, University of Innsbruck, A-6020 Innsbruck, Austria}

\author{Roeland ter Hoeven}
\affiliation{Parity Quantum Computing GmbH, A-6020 Innsbruck, Austria}
\affiliation{Institute for Theoretical Physics, University of Innsbruck, A-6020 Innsbruck, Austria}

\author{Benjamin E.~Niehoff}
\affiliation{Parity Quantum Computing GmbH, A-6020 Innsbruck, Austria}

\author{Maike Drieb-Sch\"on}
\affiliation{Parity Quantum Computing GmbH, A-6020 Innsbruck, Austria}
\affiliation{Institute for Theoretical Physics, University of Innsbruck, A-6020 Innsbruck, Austria}

\author{Wolfgang Lechner}
\email{wolfgang@parityqc.com,\\wolfgang.lechner@uibk.ac.at}
\affiliation{Parity Quantum Computing GmbH, A-6020 Innsbruck, Austria}
\affiliation{Institute for Theoretical Physics, University of Innsbruck, A-6020 Innsbruck, Austria}
\thanks{\\K.~Ender, R.~ter Hoeven and B.E.~Niehoff contributed equally.}

\begin{abstract}

We introduce parity quantum optimization with the aim of solving optimization problems consisting of arbitrary $k$-body interactions and side conditions using planar quantum chip architectures. The method introduces a decomposition of the problem graph with arbitrary $k$-body terms using generalized closed cycles of a hypergraph. Side conditions of the optimization problem in form of hard constraints can be included as  open cycles containing the terms involved in the side conditions. The generalized parity mapping thus circumvents the need to translate optimization problems to a quadratic unconstrained binary optimization problem (QUBO) and allows for the direct encoding of higher-order constrained binary optimization problems (HCBO) on a square lattice and full parallelizability of gates.   
 
\end{abstract}
\maketitle

\section{Introduction}
Hard optimization problems are omnipresent in all fields of science, technology, and industry \cite{kirkpatrick1983optimization,np_problems}. Classical algorithms for solving optimization problems have been studied and refined over decades and are now at a mature stage where fundamental improvements are unlikely. An answer to the pressing need for faster and more efficient optimization algorithms may be provided by quantum devices. The impressive recent progress in building highly coherent quantum computers based on ions \cite{cirac1995quantum,blatt2012quantum,kielpinski2002architecture}, atoms \cite{jaksch2000fast,henriet2020quantum,saffman2010quantum,bloch2008many,bernien2017probing}, superconducting circuits \cite{koch2007charge,wallraff2004strong,johnson2011quantum,arute2019quantum}, crystal defects \cite{childress2006coherent}, and photonic systems \cite{o2009photonic,qiang2018large} has led to a considerable interest in quantum algorithms to solve hard optimization problems. Quantum optimization algorithms, be they digital \cite{qaoa_speedup, qaoa_locality} or analog \cite{albash2018adiabatic,kadowaki1998quantum,annealing_methods}, employ radically new paradigms of computation.  Their basic principle is to encode an optimization problem as a physical quantum system whose energy corresponds to the problem's cost function. Solving the optimization problem is then equivalent to finding the lowest-energy configuration of the physical system.  Recently, quantum algorithms for solving optimization problems have been developed for both analog and digital quantum computers \cite{electronic_structure, protein_folding, factorizing, financial_crash}. Their possible speedup compared to classical algorithms, scaling and robustness to error, however, are open questions.

Here, we present a novel abstraction of optimization problems based on the parity transformation. This transformation maps a subset of $k$ spins into a single parity qubit whose value is equal to their $k$-fold product.  With this transformation, an optimization problem of arbitrary connectivity can be encoded in a physical system with only \emph{local} connectivity.  The parameters of the optimization problem become local magnetic fields, and the consistency conditions of the mapping become 3- or 4-body physical interactions whose strengths are independent of the logical problem.  For a wide variety of problems, it is then possible to lay out the parity-transformed qubits in a 2-dimensional grid on a physical device such that all the necessary couplings can be implemented.  The technique can just as well encode constrained optimization problems with various types of constraints.  We present the mathematical foundations of the parity transformation, which allows one to encode optimization problems with arbitrary $k$-body interaction terms onto a realistic quantum computing device, along with examples of device layouts obtained by the compiler employing this transformation.

We consider optimization problems represented by a Hamiltonian of the form 
\begin{align}\label{H_general}
H &= \sum_{i=1}^N J_i \sigma_z^{(i)} + \sum_{i=1}^N \sum_{j>i} J_{ij} \sigma_z^{(i)} \sigma_z^{(j)} \\ \nonumber 
&+ \sum_{i=1}^N \sum_{j>i} \sum_{k>j} J_{ijk} \sigma_z^{(i)} \sigma_z^{(j)} \sigma_z^{(k)} + \ldots,
\end{align}
which contains arbitrary higher-order $k$-body spin terms. In the current state-of-the-art methods, these higher-order $k$-body terms would have to be decomposed into quadratic terms using (a potentially large number of) auxiliary qubits. Our proposal instead allows $k$-body terms to be directly implemented in the physical layout, omitting the need for the intermediate mapping to quadratic unconstrained binary optimization (QUBO).

In addition to this Hamiltonian, we consider imposing constraints of the form
\begin{align}
C(\{\sigma_z^{(i)}\}) &= \delta_0 + \sum_{i=1}^N c_i \sigma_z^{(i)} + \sum_{i=1}^N \sum_{j>i} c_{ij} \sigma_z^{(i)} \sigma_z^{(j)} \\ \nonumber 
&+ \sum_{i=1}^N \sum_{j>i} \sum_{k>j} c_{ijk} \sigma_z^{(i)} \sigma_z^{(j)} \sigma_z^{(k)} + \ldots,
\end{align}
where one now seeks to optimize over the constrained subspace of states satisfying $C(\{\sigma_z^{(i)}\}) \ket{\psi} = 0$.  These constraints amount to arbitrary polynomials of the spin variables $C(\{s_z^{(i)}\}) = 0$ for $s_z^{(i)} = \pm 1$ (provided that the coefficients $\delta_0$, $c_i$, $c_{ij}$, $c_{ijk}, \ldots$ are consistent with the existence of solutions).  This covers the cases of product constraints and sum constraints, as well as more general constraints \cite{constraintpaper}.  

In this paper we introduce the parity transformation including problems with product constraints. In the associated publication \cite{constraintpaper} we describe the implementation of general constraints and in Ref.~\cite{benchmarkpaper} we benchmark the resource requirements for the quantum approximate optimization algorithm in the parity architecture for various classes of problems.

\section{The Parity Transformation}
The parity transformation is the generalization of the LHZ mapping \cite{Lechnere1500838} for hypergraphs and side conditions in the optimization problem. Let us review the LHZ mapping for completeness: the mapping is a representation of all-to-all graphs with $k$-body interactions using parity variables. For all-to-all pair interactions between $N$ logical qubits, the LHZ representation is a 2D square lattice with 4-body interactions and $N(N-1)/2$ physical qubits. The 4-body interactions are derived from closed cycles in the logical graph. The condition of closed cycles can be understood from the following considerations: the physical qubits $\tilde \sigma_z^{(ij)} = \sigma_z^{(i)}\sigma_z^{(j)}$  represent the product of two logical qubits $i$ and $j$. The product of four such physical qubits in a closed cycle is thus of the form $\sigma_z^{(i)}\sigma_z^{(j)}\sigma_z^{(j)}\sigma_z^{(k)}\sigma_z^{(k)}\sigma_z^{(l)}\sigma_z^{(l)}\sigma_z^{(i)} = 1$, which always comes out to $+1$ as all indices must appear twice.

In \cite{Lechnere1500838} a slightly generalized LHZ construction is also presented, which maps a 3-body all-to-all graph onto a 3D cubic lattice, with 4-body cyclic constraints arranged on its faces.  In principle such a construction can be generalized to $k$-body all-to-all problems, putting them on a $k$-dimensional $k$-cubic lattice, although this is impractical for realistic devices.  In general such constructions require a number of lattice qubits scaling as $N^k$; for 2D devices, this means one can represent at best 2-body all-to-all problems, using $O(N^2)$ physical qubits.
We present here a further generalization, called the `parity transformation', that can represent \emph{arbitrary} optimization problems (that is, with mixtures of $k$-body terms of various $k$) on a 2D square lattice (that is, without resorting to higher-dimensional constructions).  The required number of physical qubits scales as $K$, where $K$ is the number of non-zero terms in the Hamiltonian.  The same mapping can also encode constrained optimization problems without overhead.  There is no need to rewrite the Hamiltonian using QUBO-like transformations, as the $k$-body interactions are encoded directly.
\begin{figure}[tb!]
\centering
\includegraphics{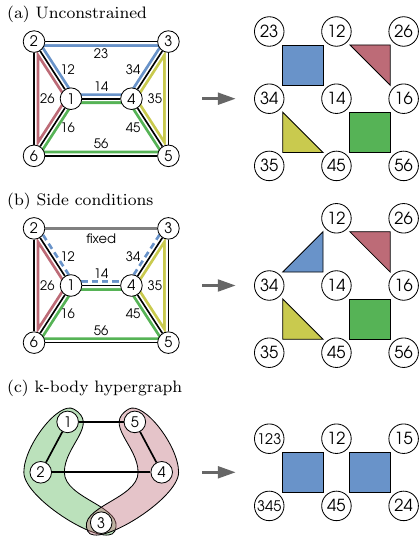}
\caption{(a) An unconstrained optimization problem is represented by a graph consisting of spins (circles) and edges (black lines). The parity construction is based on finding closed loops of length 3 and 4, e.g., the red triangle connecting spins 1,2 and 6 is a closed loop of length 3. The compiler constructs parity constraints from the closed loops and connects them to a compact quantum chip. In the illustration, colors of the constraints (right) match the colors of the closed loops (left). (b) A side condition to an optimization problem is a hard constraint that has to be satisfied in the solution. As an illustration, we consider the same problem as in (a) but add the side condition that $\sigma_z^{(2)}\sigma_z^{(3)} = +1$. In the parity picture, this can be encoded in \textit{open loops} (blue dashed line). The loop of length 4 is translated by the compiler to a parity constraint between 3 qubits (blue triangle). (c) An optimization problem with higher-order terms is encoded in hypergraphs, e.g., a term $\sigma_z^{(1)}\sigma_z^{(2)}\sigma_z^{(3)}$ is indicated by the green shape. In the parity transformation, higher order terms can be treated exactly the same as edges and thus higher-order terms are encoded without decomposition into a QUBO problem first.}
\label{fig:examples}
\end{figure}
\subsection{The embedding and layout maps}
It is convenient to think of the transformation, from logical problem to chip layout, as the composition of two maps, which we will call the \emph{embedding} and \emph{layout} maps.  These are maps between \emph{optimization problems}, and thus must preserve the algebra of Pauli $Z$ operators, while mapping states between the following Hilbert spaces:
\begin{equation} \label{P L def}
	\cH_\text{log} \xhookrightarrow{\cP} \cH_\text{par} \xrightarrow{\mathscr{L}} \cH_\text{phys}.
\end{equation}
The embedding map $\cP$ is a linear map that represents every $k$-fold product of logical qubits as a single parity qubit, and takes the logical Hilbert space $\cH_\text{log}$ to a code subspace in the `parity Hilbert space' $\cH_\text{par}$ spanned by those parity qubits.  The code subspace can be obtained from a set of consistency conditions acting as projections, which arise from demanding that the algebra of Pauli $Z$ operators be preserved. The layout map $\cL$ then places the parity qubits on a physical device by mapping them 1-to-1 onto the physical Hilbert space $\cH_\text{phys}$ in such a way that the consistency conditions can be realized as local interactions. The first map $\cP: \cH_\text{log} \hookrightarrow \cH_\text{par}$ is straightforward and can be expressed essentially in closed form. The second map $\cL: \cH_\text{par} \rightarrow \cH_\text{phys}$ is hard to compute and typically requires searching the space of possible qubit layouts.  In this section we focus on how the embedding map $\cP$ is constructed.

\subsection{Construction of the parity map}

We first define the parity Hilbert space $\cH_\text{par}$.  Suppose the logical Hilbert space $\cH_\text{log}$ consists of $N$ qubits.  The parity transformation will represent each interaction term of the Hamiltonian as a single parity qubit.  There are $K$ such interaction terms, and thus $\cH_\text{par}$ is a tensor product of $K$ qubits for some $0 < K < 2^N$, depending on which $k$-body interactions are turned on.  We denote Pauli $Z$ operators on the $i$-th logical qubit by $\sigma_z^{(i)}$ for $i \in 1 \ldots N$, and on the $a$-th parity qubit by $\tilde \sigma_z^{(a)}$ for $a \in 1 \ldots K$.  The parity transformation is then given by a collection of mapping relations of the form
\begin{equation} \label{parity map}
	\sigma_z^{(i_1)} \sigma_z^{(i_2)} \ldots \sigma_z^{(i_k)} \mapsto \tilde \sigma_z^{(a)},
\end{equation}
which maps the $k$-fold product of logical qubits $i_1, i_2, \ldots i_k$ onto a single qubit $a$ in $\cH_\text{par}$.  However, the $\tilde \sigma_z^{(a)}$ are a new set of Pauli $Z$ operators and do not respect the algebra implied by the original $\sigma_z^{(i)}$.  In order to preserve the Pauli $Z$ algebra, one must impose a set of consistency conditions, which can be obtained from cyclic products of the $\sigma_z^{(i)}$ of the form
\begin{equation} \label{cyclic condition}
	\sigma_z^{(i_1)} \sigma_z^{(i_2)} \ldots \sigma_z^{(i_m)} = 1,
\end{equation}
where each logical qubit index appears an even number of times. Such cyclic conditions are generalized closed cycles, that correspond to cycles in the logical graph in the original LHZ construction.  In both the original case and the generic case with $k$-body interactions, one obtains \emph{parity projection conditions} on $\cH_\text{par}$,
\begin{equation} \label{projection condition}
	\tilde \sigma_z^{(a_1)} \tilde \sigma_z^{(a_2)} \ldots \tilde \sigma_z^{(a_n)} \ket{\tilde \psi} = \ket{\tilde \psi},
\end{equation}
for every combination of parity qubits whose pre-image along the parity map \eqref{parity map} would give a cyclic product \eqref{cyclic condition}.  The Pauli $Z$ algebra of $\cH_\text{log}$ is then preserved along the subspace of $\cH_\text{par}$ satisfying \eqref{projection condition}, which is effectively the code subspace.
Let us illustrate this using the example in Fig.~\ref{fig:examples}(c), given by the Hamiltonian
\begin{equation} \label{example hamiltonian}
	\begin{split}
		H &= J_{12} \, \sigma_z^{(1)} \sigma_z^{(2)} + J_{15} \, \sigma_z^{(1)} \sigma_z^{(5)} \\
		&+ J_{24} \, \sigma_z^{(2)} \sigma_z^{(4)} + J_{45} \, \sigma_z^{(4)} \sigma_z^{(5)} \\
		&+ J_{123} \, \sigma_z^{(1)} \sigma_z^{(2)} \sigma_z^{(3)} + J_{345} \, \sigma_z^{(3)} \sigma_z^{(4)} \sigma_z^{(5)}.
	\end{split}
\end{equation}
If we label the interaction terms by the combination of logical qubits to which they correspond, then we can write the parity transformation as
\begin{align}
	\sigma_z^{(1)} \sigma_z^{(2)} &\mapsto \tilde \sigma_z^{(1, 2)}, &
	\sigma_z^{(1)} \sigma_z^{(5)} &\mapsto \tilde \sigma_z^{(1, 5)}, \notag \\
	\sigma_z^{(2)} \sigma_z^{(4)} &\mapsto \tilde \sigma_z^{(2, 4)}, &
	\sigma_z^{(4)} \sigma_z^{(5)} &\mapsto \tilde \sigma_z^{(4, 5)}, \label{example parity map} \\
	\sigma_z^{(1)} \sigma_z^{(2)} \sigma_z^{(3)} &\mapsto \tilde \sigma_z^{(1, 2, 3)}, &
	\sigma_z^{(3)} \sigma_z^{(4)} \sigma_z^{(5)} &\mapsto \tilde \sigma_z^{(3, 4, 5)}, \notag
\end{align}
and one can observe that certain cyclic conditions must hold on the code subspace, such as
\begin{equation} \label{cycle projection}
	\tilde \sigma_z^{(1, 2)} \tilde \sigma_z^{(4, 5)} \tilde \sigma_z^{(1, 2, 3)} \tilde \sigma_z^{(3, 4, 5)} \ket{\tilde \psi} = \ket{\tilde \psi},
\end{equation}
which corresponds, in the logical qubits, to the product
\begin{equation}
	\sigma_z^{(1)} \sigma_z^{(2)} \sigma_z^{(4)} \sigma_z^{(5)} \sigma_z^{(1)} \sigma_z^{(2)} \sigma_z^{(3)} \sigma_z^{(3)} \sigma_z^{(4)} \sigma_z^{(5)} = 1.
\end{equation}
One can in principle search for such cyclic products and construct all possible projection conditions in order to define the code subspace in $\cH_\text{par}$.  Once the parity map and projection conditions are constructed, the layout map $\cL : \cH_\text{par} \to \cH_\text{phys}$ must place all of the parity qubits $\tilde \sigma_z^{(a)}$ on the physical device in such a way that the projection conditions in \eqref{projection condition} can be realized.  In Fig.~\ref{fig:examples} (c), one can see that the projection \eqref{cycle projection} has been realized as a square plaquette.  However, rather than conducting a search through the logical (hyper)graph to find cyclic products, we find it useful to take a more algebraic approach using the language of classical linear codes.  The parity map and projection conditions can be packaged into a \emph{generator matrix} $\mathbf{G}$ and a \emph{parity check matrix} $\mathbf{P}$, which can be manipulated using standard tools of linear algebra, as we now describe.

\subsection{Generator and parity check matrices}

For an optimization problem with $N$ logical qubits and $K$ interaction terms in the Hamiltonian (whether they be single-body, two-body, or generic $k$-body interactions), one can describe the code subspace in $\cH_\text{par}$ via an $N \times K$ \emph{generator matrix} $\mathbf{G}$ which maps logical bit-string  $\mathbf{v}$ into encoded bit-string $\mathbf{w}$ via right-multiplication:
\begin{equation} \label{G definition}
	\mathbf{w} = \mathbf{v} \mathbf{G}.
\end{equation}
All linear algebra here should be understood modulo 2; in the bit-strings $\mathbf{w}$ and $\mathbf{v}$, a 0 represents the spin eigenstate $\ket{0}$ (with $\sigma_z$ eigenvalue $+1$), whereas a 1 represents $\ket{1}$ (with $\sigma_z$ eigenvalue $-1$).  To construct $\mathbf{G}$ from the mapping in \eqref{parity map}, let the $a$-th column have a 1 in positions $i_1, i_2, \ldots i_k$, and a $0$ elsewhere.  In the example problem \eqref{example hamiltonian}, one has $N = 5$ logical qubits and $K = 6$ interaction terms.  Each interaction term corresponds to a column of $\mathbf{G}$, which gives us:
\begin{equation} \label{example generator matrix}
	\mathbf{G} = \begin{pmatrix}
		1 & 1 & 0 & 0 & 1 & 0 \\
		1 & 0 & 1 & 0 & 1 & 0 \\
		0 & 0 & 0 & 0 & 1 & 1 \\
		0 & 0 & 1 & 1 & 0 & 1 \\
		0 & 1 & 0 & 1 & 0 & 1
	\end{pmatrix}.
\end{equation}
Given the generator matrix, we must then construct the \emph{parity check matrix} $\mathbf{P}$ which effectively encapsulates all the projection conditions. The code subspace in $\cH_\text{par}$ is defined by the set of all bit-strings $\mathbf{w}$ which satisfy
\begin{equation} \label{P definition}
	\mathbf{w} \mathbf{P}^\top = 0.
\end{equation}
Each row of $\mathbf{P}$ represents a single projection condition acting on $\cH_\text{par}$.  One can determine candidate rows of $\mathbf{P}$ by searching for combinations of columns of $\mathbf{G}$ which sum to 0 modulo 2; a row then has a 1 in the positions of the columns participating in a sum, and a 0 elsewhere.  From the generator matrix in \eqref{example generator matrix}, we can determine
\begin{equation} \label{example parity matrix}
	\mathbf{P} = \begin{pmatrix}
		1 & 1 & 1 & 1 & 0 & 0 \\
		1 & 0 & 0 & 1 & 1 & 1
	\end{pmatrix}.
\end{equation}
More generally, the check matrix is the matrix of maximal rank which satisfies
\begin{equation}
	\mathbf{G} \mathbf{P}^\top = 0,
\end{equation}
that is, the rows of $\mathbf{P}$ constitute a basis of the null space of $\mathbf{G}$.  Any such basis gives a valid $\mathbf{P}$, and all such bases are related by elementary row operations.  The number of rows of $\mathbf{P}$ is always at least $K - N$.  In this example, it is $K - N + 1$ because there is a degeneracy in the Hamiltonian \eqref{example hamiltonian} (it is invariant under the simultaneous spin flip of qubits 1, 2, 4, 5).  For a Hamiltonian with $D$ degeneracies, the number of rows of $\mathbf{P}$ is $K - N + D$. 

The check matrix $\mathbf{P}$ gives a complete basis of projection conditions needed to define the code subspace of $\cH_\text{par}$ on which the mapping relations \eqref{parity map} are consistent.  Each projection condition must be realized on the physical device as a constraint coupling, and it will be the task of the layout mapping $\cL : \cH_\text{par} \to \cH_\text{phys}$ to arrange the physical qubits in such a way that every projection condition can be realized (either by a 3- or 4-body plaquette coupling, or a chain of CNOTs).

\subsection{Readout and decoding}
Having mapped the original optimization problem onto the set of parity qubits, one must be able to undo this map in order to interpret solutions. Let us first consider the case where all constraints are satisfied. The $N \times K$ \emph{decoding matrix} $\mathbf{D}$ maps encoded bit-strings $\mathbf{w}$ back into logical bit-strings $\mathbf{v}$:
\begin{equation} \label{D definition}
	\mathbf{w} \mathbf{D}^\top = \mathbf{v}.
\end{equation}
To find $\mathbf{D}$, we first observe that valid encoded bit-strings always satisfy the projection $\mathbf{w} = \mathbf{w} (1 - \mathbf{P}^\top \mathbf{P})$ as a consequence of \eqref{P definition}.  Acting with $\mathbf{G}$ on \eqref{D definition} from the right, and using \eqref{G definition} to eliminate $\mathbf{w}$, we obtain
\begin{equation} \label{D equation}
	(1 - \mathbf{G} \mathbf{D}^\top) \mathbf{G} (1 - \mathbf{P}^\top \mathbf{P}) = 0,
\end{equation}
which means that $\mathbf{D}^\top$ is a right-pseudoinverse of $\mathbf{G}$, on the subspace projected out by $\mathbf{P}$.  That is, given any $\mathbf{D}$ which solves \eqref{D equation}, one can obtain another valid solution by adding any row of $\mathbf{P}$ to any row of $\mathbf{D}$.  Any such $\mathbf{D}$ is a valid decoding matrix.  For example, corresponding to the generator matrix in \eqref{example generator matrix}, one can find
\begin{equation} \label{example decoding matrix}
	\mathbf{D} = \begin{pmatrix}
		1 & 0 & 1 & 1 & 0 & 0 \\
		0 & 0 & 1 & 1 & 0 & 0 \\
		0 & 0 & 0 & 1 & 0 & 1 \\
		0 & 0 & 0 & 1 & 0 & 0 \\
		0 & 0 & 0 & 0 & 0 & 0
	\end{pmatrix};
\end{equation}
however, another equivalent solution is
\begin{equation}
	\mathbf{D} = \begin{pmatrix}
		0 & 0 & 1 & 0 & 1 & 1 \\
		0 & 0 & 1 & 1 & 0 & 0 \\
		0 & 0 & 0 & 1 & 0 & 1 \\
		0 & 0 & 0 & 1 & 0 & 0 \\
		0 & 0 & 0 & 0 & 0 & 0
	\end{pmatrix},
\end{equation}
which can be obtained by adding the second row of \eqref{example parity matrix} to the first row of \eqref{example decoding matrix}.  They are equivalent decoding matrices because of the algebraic relationships satisfied by valid encoded bit-strings in \eqref{P definition}.

The decoding matrix $\mathbf{D}$ informs us how to interpret any state in $\cH_\text{par}$ which lies in the code subspace defined by the projection conditions.  However, in experimental implementations of annealing and digital optimization processes, the quantum state of the physical system may be driven out of the code subspace (either explicitly by the driver Hamiltonian, or by the accumulation of errors), and thus some error-correction scheme must be used before $\mathbf{D}$ can be applied.  In this case, the non-uniqueness of $\mathbf{D}$ becomes important, as it implies that there is no uniquely `correct' logical configuration corresponding to a state in $\cH_\text{par}$ that fails to satisfy the projections \eqref{projection condition}.  The error-correction process is effectively a method of deciding how to project a generic state in $\cH_\text{par}$ onto the code subspace, where $\mathbf{D}$ can be applied.

One useful method of making such decisions is belief propagation, as proposed in Ref.~\cite{pastawski2016error}.  An alternative (perhaps less useful) way is to read out only a minimal subset of parity qubits which are sufficient to reconstruct a logical state, effectively ignoring parity checks.

A simple error-correction scheme was proposed in \cite{Lechnere1500838} that builds upon this `minimal readout' strategy to construct a more robust model.  Instead of taking \emph{one} minimal readout, several different ones are taken, on different subsets of qubits, and given a majority vote to determine the most likely logical state.  In the original LHZ construction (for 2-body all-to-all problems), a minimal readout subset corresponds precisely to a spanning tree in the logical graph.  Thus one can either enumerate \emph{all} such spanning trees and take this majority vote, or draw only a limited number of them from some distribution.

Although it is difficult to generalize the notion of a `spanning tree' to the hypergraph that represents a problem with $k$-body interactions, it is straightforward to generalize this notion of `majority vote between minimal readout subsets'.  Once the full \emph{layout} is known (for example, in Fig.~\ref{plaquette example}), one can construct a minimal readout set as follows:  Begin at any arbitrary plaquette in the layout.  Assign all but one of its qubits to the minimal readout set, and mark them `read out'.  Since these qubits are to be read out, their values are known.  The remaining qubit in this plaquette is unknown, but we assume the constraint represented by this plaquette holds, and thus the value of this qubit can be deduced from the read out qubits; mark it `determined'.  To continue constructing the readout set, choose any plaquette which is adjacent to those plaquettes already visited; that plaquette will have some qubits `read out', some qubits `determined', and some qubits unmarked.  Of the unmarked qubits, mark all but one of them as `read out', and the final one as `determined'.  Continue in this fashion until all qubits have been marked.  Then the ones marked `read out' constitute a minimal readout subset.

\section{Constrained optimization problems}

For some optimization problems, one may wish to impose additional conditions on the possible solutions, beyond the Hamiltonian \eqref{H_general}.  Such \emph{constrained optimization problems} have a number of constraints which act in the logical spin space $\cH_\text{log}$.  These can be product constraints $\sigma_1 \sigma_2 \sigma_3 = 1$ or sum constraints $\sigma_1 + \sigma_2 + \sigma_3 = 1$ (or generically, polynomial constraints, which combine both principles).

\subsection{Product constraints}
Product constraints can be realized as projections in the logical Hilbert space $\cH_\text{log}$.  A collection of constraints of the form
\begin{equation}
	\sigma_{i_1} \sigma_{i_2} \cdots \sigma_{i_k} \ket{\psi} = \ket{\psi}
\end{equation}
can be represented, in our linear-algebra language, by imposing a condition on logical bit-strings:
\begin{equation}
	\mathbf{v} \mathbf{C}^\top = 0,
\end{equation}
where the \emph{constraint matrix} $\mathbf{C}$ has $N$ columns and any number of rows; in a given row, a 1 in the $i$-th column indicates the presence of $\sigma_i$ in a product constraint.  The rest of the algebraic structure of the linear code goes through as described above, but with the caveat that the space of logical bit-strings has been restricted to a subspace in $\cH_\text{log}$ given by the projection
\begin{equation}
	\mathbf{v} (1 - \mathbf{C}^\top \mathbf{C}) = \mathbf{v}.
\end{equation}
The parity check matrix $\mathbf{P}$ is then given by the matrix of maximal rank which satisfies
\begin{equation}
	(1 - \mathbf{C}^\top \mathbf{C}) \mathbf{G} \mathbf{P}^\top = 0,
\end{equation}
and the decoding matrix $\mathbf{D}$ is any simultaneous solution of
\begin{align}
	(1 - \mathbf{C}^\top \mathbf{C}) (1 - \mathbf{G} \mathbf{D}^\top) \mathbf{G} (1 - \mathbf{P}^\top \mathbf{P}) &= 0, \\
	(1 - \mathbf{P}^\top \mathbf{P}) \mathbf{D}^\top \mathbf{C}^\top &= 0.
\end{align}
That is, $\mathbf{D}^\top$ is a right-pseudoinverse of $\mathbf{G}$ on the constrained subspace of logical bit-strings, and also satisfies the constraint condition itself for all valid encoded bit-strings $\mathbf{w} = \mathbf{w} (1 - \mathbf{P}^\top \mathbf{P})$.

For our example Hamiltonian \eqref{example hamiltonian}, suppose we impose the constraint $\sigma_1 \sigma_2 \sigma_4 = 1$, which is sufficient to fix the spin-flip degeneracy in qubits $1, 2, 4, 5$.  This is represented by the constraint matrix
\begin{equation}
	\mathbf{C} = \begin{pmatrix}
		1 & 1 & 0 & 1 & 0
	\end{pmatrix}.
\end{equation}
The parity check matrix $\mathbf{P}$ is unchanged, since this constraint only fixes a degeneracy and has no effect on the code subspace in $\cH_\text{par}$.  The decoding matrix becomes
\begin{equation}
	\mathbf{D} = \begin{pmatrix}
		0 & 0 & 1 & 0 & 0 & 0 \\
		1 & 0 & 1 & 0 & 0 & 0 \\
		0 & 0 & 0 & 1 & 0 & 1 \\
		1 & 0 & 0 & 0 & 0 & 0 \\
		1 & 0 & 0 & 1 & 0 & 0
	\end{pmatrix},
\end{equation}
which can be obtained from \eqref{example decoding matrix} via elementary row operations. In Ref.~\cite{constraintpaper} we show how to encode the more general sum constraints.

\section{Compilation}
\label{sec:compilation}

Having defined the mapping from logical qubits to qubits in parity space $\cH_\text{par}$ and obtained the necessary basis of projection conditions encoded in the parity matrix $\mathbf{P}$, we now discuss the second step of the compilation process, the layout map $\cL : \cH_\text{par} \to \cH_\text{phys}$, which implements the various code subspace projections on a physical device.  The qubits of $\cH_\text{par}$ must be mapped 1-to-1 onto physical qubits, in such a way that the projection conditions $\mathbf{P}$ can be realized by physical couplings.  The physical qubits and available couplings are at fixed locations on the device, and thus computing a valid layout $\cL$ is a problem akin to graph isomorphism, and requires some sort of search algorithm.  In this section we will discuss some techniques by which the parity compiler achieves such layouts.

There are two main types of devices we consider, based on how the projection conditions of $\mathbf{P}$ can be realized: \emph{plaquette devices} and \emph{CNOT devices}.  

\subsection{Compilation via plaquettes}
\begin{figure}[t]
\centering
	\includegraphics{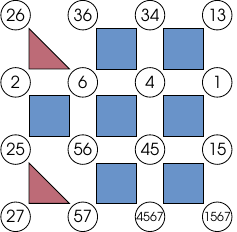}
	\caption{Example of a plaquette compilation from a logical Hamiltonian on 7 logical qubits with 16 interactions. Because of the four-body interactions, this problem cannot be directly encoded using the LHZ mapping. The physical qubits are labeled with the combination of logical qubits that participate in the corresponding interaction.  The code subspace is obtained from a combination of projectors of length 4 (blue) and length 3 (red).}
	\label{plaquette example}
\end{figure}
We will first consider plaquette compilation.  On a plaquette device there exist a number of 3- and 4-body couplers which can constrain the parity of spins situated in a square or triangle.  The ideal device which we consider here has a 4-body coupler available on every grid square, which can be (optionally) chosen to constrain only 3 corners of the square, making it serve as a 3-body coupler when needed.

Plaquette compilation of Hamiltonians with higher-order $k$-body terms (HCBO) as in \eqref{H_general} is a straightforward generalization of the original LHZ proposal for Hamiltonians with 2-body interactions.  The projections onto the code subspace are realized by square or triangle couplers in a square lattice, as in Fig.~\ref{plaquette example}.  These couplers add terms to the physical Hamiltonian of the form
\begin{equation}
	- C_{ijk(\ell)} \, \tilde \sigma_z^{(i)} \tilde \sigma_z^{(j)} \tilde \sigma_z^{(k)} (\tilde \sigma_z^{(\ell)}),
\end{equation}
where $C$ is some coupling strength chosen to be sufficiently large, so that the physical ground state is one where the product of spins going around every plaquette is $+1$.

Using plaquettes on a square lattice, it is only possible to realize projectors with 3 or 4 $\sigma_z$ operators.  This means that the rows of the parity check matrix $\mathbf{P}$ must have Hamming weight 3 or 4.  If this is not the case, then there are a few options.  First, one may use elementary row operations on $\mathbf{P}$ to transform it into a matrix whose rows have Hamming weight 3 or 4.  If this is not possible, then one can instead introduce ancilla qubits in order to decompose longer projectors into shorter ones.  For example, the projector
\begin{equation}
	\tilde \sigma_z^{(1)} \tilde \sigma_z^{(2)} \tilde \sigma_z^{(3)} \tilde \sigma_z^{(4)} \tilde \sigma_z^{(5)} = 1
\end{equation}
can be realized as the combination of two projectors
\begin{equation}
	\tilde \sigma_z^{(1)} \tilde \sigma_z^{(2)} \tilde \sigma_z^{(6)} = 1, \qquad \sigma_z^{(3)} \tilde \sigma_z^{(4)} \tilde \sigma_z^{(5)} \tilde \sigma_z^{(6)} = 1,
\end{equation}
where 6 is a (physical) ancilla qubit.

\subsubsection{Implementing side conditions}
\begin{figure}[bt]
\centering
	\includegraphics{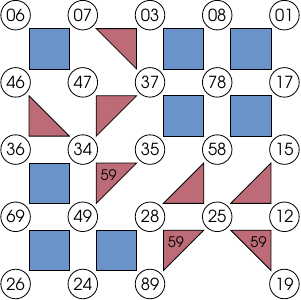}
	\caption{Example plaquette compilation with side conditions. The original Hamiltonian has 10 logical qubits appearing in 24 interactions, which requires 45 physical qubits in the LHZ mapping. The physical qubits are labeled by the combination of logical qubits in the corresponding interaction term.  The constraint $\sigma_z^{(5)} \sigma_z^{(9)} = 1$ is imposed in the logical problem.  Elementary row operations on $\mathbf{P}$ have been used to subsume this constraint into three projection conditions, $\tilde \sigma_z^{(3, 4)} \tilde \sigma_z^{(3, 5)} \tilde \sigma_z^{(4, 9)} = 1$, $\tilde \sigma_z^{(2, 8)} \tilde \sigma_z^{(2, 5)} \tilde \sigma_z^{(8, 9)} = 1$, and $\tilde \sigma_z^{(2, 5)} \tilde \sigma_z^{(1, 2)} \tilde \sigma_z^{(1, 9)} = 1$.}
	\label{fixed_inters}
\end{figure}

We have discussed how constrained optimization problems with product constraints can be subsumed into the same linear-algebraic framework for parity compilation.  Essentially the parity check matrix $\mathbf{P}$ is constructed such that it contains all the necessary information already.  In Fig.~\ref{fixed_inters} we give an example.

\subsubsection{Choosing ancillas}
Occasionally it is necessary to add ancilla qubits in order to compile certain problems.  An ancilla qubit in this context is any physical qubit which does not correspond to a term in the original Hamiltonian; its only purpose is to assist us in implementing the parity projections on the other qubits.

\begin{figure}[t]
\centering
	\includegraphics[]{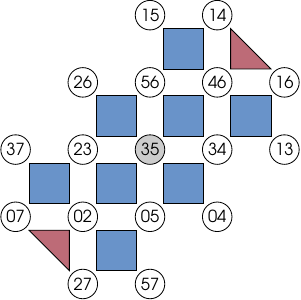}
	\caption{Plaquette compilation using a pre-constructed ancilla, shown in grey. The ancilla qubit $\sigma_z^{(3, 5)}$ makes it possible to lay out the constraints. Note that here 17 physical qubits are required compared to the 28 needed in the LHZ mapping. 
	}
	\label{pre_ancilla}
\end{figure}

There are essentially two situations in which an ancilla may need to be added.  The first situation is one we have already mentioned: when the rows of $\mathbf{P}$ have Hamming weight greater than 4, and this cannot be remedied by elementary row operations.  In this case, an ancilla must be created to break up a long projector into shorter ones.  One can make choices about exactly how to do this, since a given long projector might be broken up in many different ways.  In some cases, there may be heuristics which prefer one choice over another, such as when a single ancilla may serve to break up several long projectors.  An example of a compiled problem using an ancilla qubit is given in Fig.~\ref{pre_ancilla}.

The second situation for using ancillas is for additional flexibility in the search for a valid layout.  For example, one may choose, dynamically during the course of compilation, to break up a length-4 projector by adding an ancilla which splits it into two length-3 projectors.  The advantage is that the original 4 physical qubits no longer have to be adjacent in the layout.  Such `dynamical ancillas' allow one to find layouts that might otherwise be impossible. 

In general, the main drawback of adding ancillas is that for every ancilla, one also must add an extra projection (effectively, the matrix $\mathbf{P}$ grows by one row).  So more physical resources (qubits and couplings) are required for a given logical problem size.  However, sometimes such ancillas are necessary in order to find a layout for a given problem.  They must be chosen judiciously.

\subsection{Compilation for digital optimization}

Next we will consider compilation to devices whose couplings are CNOT gates acting along the edges of a square lattice, and local $Z$ rotations acting on the vertices.  These gates can be used to implement unitary operators corresponding to time evolution under spin parity constraints, suitable for digital optimization algorithms \cite{farhi2014quantum, Lechner2018, Moll_2018}.  The projection conditions given by the parity matrix $\mathbf{P}$ can be realized with these CNOT gates and $Z$ rotations rather than with 4-body plaquette couplers, as we will describe.  This type of compilation is well-suited to devices with an array of CNOT gates (or equivalent), and it allows additional flexibility in constructing the physical layout, which provides further opportunity for optimizations, such as increasing parallelization or minimizing circuit depth.

\begin{figure}[t]
\centering
	\includegraphics[]{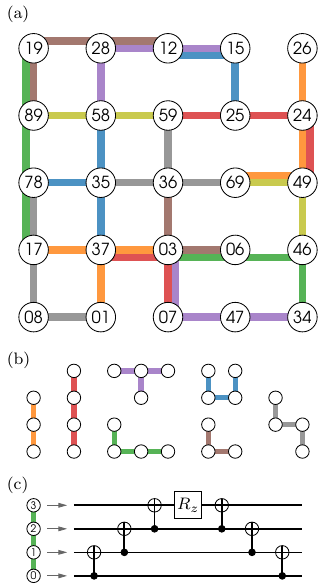}
	\caption{(a) Example of a compilation implementing the code space projectors with CNOT gates. Each coloured path represents a single projector which constrains the physical qubits along that path to have even parity. Implementing this using the LHZ mapping requires 45 physical qubits. (b) All possible shapes of projectors used during the layout process. (c) The decomposition of a projection condition into a chain of CNOT gates and a rotation, implementing the unitary $U=\exp(-i\dfrac{\alpha}{2}\sigma_z^{(0)}\sigma_z^{(1)}\sigma_z^{(2)}\sigma_z^{(3)})$, where $\alpha$ is the rotation angle of $R_z$.}
	\label{line_constraints}
\end{figure}

An example of a CNOT compilation is given in Fig.~\ref{line_constraints}.  Each code space projector is realized by a tree of edges in the physical graph.  CNOT gates are executed from the leaves of the tree towards its root, tying a set of qubits together, and a local $Z$ rotation at the end of the path serves to implement the appropriate parity-constraining unitary operation. One can in principle lay out projectors of any length this way, although it is useful to bound the length in order to limit the circuit depth of the chain of CNOTs, and thus one will want to use elementary row operations on $\mathbf{P}$ or introduce ancillas, just as in the case of plaquette projectors.

The trees or paths of CNOTs may be laid out in any shape; the only requirement is that all the qubits in a given projector be contiguous in the layout.  This high degree of flexibility can greatly aid in finding such layouts.  One can also consider imposing additional conditions, such as maximizing the number of CNOTs which can be run in parallel, in order to minimize the total circuit depth.

\section{Conclusion}

We presented an architecture to encode HCBO problems i.e.\ optimization problems with higher-order interactions and side conditions to simple square lattices. This is a generalization of the original LHZ mapping in several ways: The main generalization is that from all-to-all models of pair interactions (or $k$-body terms) to problems with mixtures of $k$-body terms of different $k$ and less than all-to-all connectivity. The parity mapping translates  products of $\sigma_z$ into single parity qubits. Using the generalized conditions that replace the closed cycles in LHZ, this allows for the encoding of arbitrary mixtures of $k$-body terms. 
A main advantage of compilation compared to LHZ is the reduction of the number of qubits from an $N^2$ scaling to a $K$ scaling, where $K$ is the number of terms in the Hamiltonian, independent of the $k$-locality of these terms.
%
For digital quantum algorithms, this reduction in qubit overhead again provides an advantage in gate counts, which is amplified by the lack of need for SWAP gates, especially for dense problem graphs with higher-order interactions \cite{benchmarkpaper}.  In compiled plaquette layouts decomposed into gates for digital devices, the resulting circuits are fully parallelizable and enable algorithms to be implemented with system-size-independent circuit depth.

Another novelty reported here is that side conditions in the form of product constraints can be included without overhead. This is achieved by another generalization of the LHZ cycles, namely the generalization to open cycles. The implementation of more general constraints, i.e.\ constraints on sums of products in the logical spin variables, is investigated in \cite{constraintpaper}. Depending on the available hardware and the specific problem such constraints can also be encoded without overhead in required resources.

In combination with a compiler, i.e.\ classical software that finds the optimal layout of the plaquettes on a chip, the parity mapping is a powerful tool to encode optimization problems with side conditions on current quantum hardware. 
Taking all these advantages into account, the parity architecture may serve as a blueprint for next generation quantum optimization devices \cite{preskill2018quantum,bharti2021noisy}. 

\textit{Acknowledgements - } Work at the University of Innsbruck is supported by the European Union program Horizon 2020 under Grants Agreement No.~817482 (PASQuanS), and by the Austrian Science Fund (FWF) through a START grant under Project No. Y1067-N27 and the SFB BeyondC Project No. F7108-N38, the Hauser-Raspe foundation. This material is based upon work supported by the Defense Advanced Research Projects Agency (DARPA) under Contract No. HR001120C0068. Any opinions, findings and conclusions or recommendations expressed in this material are those of the author(s) and do not necessarily reflect the views of DARPA.


 \onecolumngrid
\end{document}